\documentclass[11pt]{article}
\usepackage{graphicx}

\newcommand{\BABARPubYear}    {02}

\newcommand{\BABARProcNumber} {024}
\newcommand{\SLACPubNumber} {9209}

\input pubboard/babarsym

\long\def\inst#1{\par\nobreak\kern 4pt\nobreak
    {\it #1}\par\vskip 10pt plus 3pt minus 3pt}

\begin{document}
{\pagestyle{empty}

\begin{flushright}
SLAC-PUB-\SLACPubNumber \\
\babar-PROC-\BABARPubYear/\BABARProcNumber \\

May, 2002 \\
\end{flushright}

\par\vskip 4cm

\begin{center}
\Large \bf Hadronic \B decays at \babar
\end{center}
\bigskip

\begin{center}
\large 
Francesco Fabozzi\\
INFN -- Sezione di Napoli \\
Complesso Universitario di Monte Sant'Angelo \\
Via Cintia, I-80126 Napoli, Italy \\
(on behalf of the \lbabar\ Collaboration)
\end{center}
\bigskip \bigskip

\begin{center}
\large \bf Abstract
\end{center}
We present preliminary results on hadronic decays of $B$ mesons, based
on data recorded at the \FourS resonance with the \babar \, detector at
the PEP-II $B$-factory at SLAC. 
We measure branching fractions of many
$B$ decay modes, including decays to $\jpsi \phi K$, $\jpsi \pip \pim$ and
$\etac K$ final states. We report the observation of the decay 
$B \ra \Ds \pim$ and the first measurement of the flavor-tagged $D$
meson production in $B^0$ decays.
Since their preliminary nature, the results presented in this paper
are based on different data samples.

\vfill
\begin{center}
Invited talk presented at the XXXVIIth Rencontres de Moriond 
on QCD and Hadronic Interactions, \\
3/16/2002---3/23/2002, Les Arcs, France
\end{center}

\vspace{1.0cm}
\begin{center}
{\em Stanford Linear Accelerator Center, Stanford University, 
Stanford, CA 94309} \\ \vspace{0.1cm}\hrule\vspace{0.1cm}
Work supported in part by Department of Energy contract DE-AC03-76SF00515.
\end{center}

\section{The \babar \, detector}

The \babar \, detector~\cite{NIMbab} at the PEP-II asymmetric-energy
$B$-factory~\cite{PEPII} at SLAC consists of a silicon vertex tracker
(SVT) for precise decay vertex determination, a 40-layer drift chamber
(DCH) for momentum and track angles measurement, a detector of
internally reflected Cherenkov radiation (DIRC) for charged hadron
identification, and a CsI(Tl) electromagnetic calorimeter (EMC) for
photon reconstruction and electron identification. 
A superconducting solenoid provides a
magnetic field of $1.5$ T, and the iron of the flux return is
instrumented with resistive plate chambers (IFR) to provide muon
identification and neutral hadron reconstruction. 

\section{Hadronic $B$ decays to charmonium}

Color suppressed transitions $b \ra c \bar{c} s (d)$ are responsible
for hadronic $B$ decays to final states containing a charmonium.
Theoretical predictions are based on the factorization hypothesis,
that can be accurately tested with extensive and precise branching
fraction determinations~\cite{jpsiPRD}.

\subsection{Rare $B$ decays to states with a $\jpsi$}

The Cabibbo-suppressed decays $B \ra \jpsi \eta (\eta^\prime)$ are
described by a $b \ra c \bar{c} d$ transition, as the
observed decay $B \ra \jpsi \pi$. 
An upper limit on the decay $B \ra \jpsi \eta$ has been
set by the L3 Collaboration~\cite{L3ref}, while there is no published 
result for the $B \ra\jpsi \eta^\prime$ channel.

The decay  $B \ra \jpsi \phi K$ is described by a $b\bar{q} \ra
c\bar{c}s\bar{s}s\bar{q}$ transition, in which the $s\bar{s}$ pair is
produced from sea quarks or via gluon emission. This mode
has been observed by the CLEO Collaboration~\cite{CLEOphiK} with
a branching fraction of ${\cal B}(B \ra \jpsi \phi K) = 
(8.8 ^{+ 3.5}_{-3.0}\pm 1.3) \times 10^{-5}$.

The decay $B \ra \jpsi \phi$, which has not yet been observed, is
explained with the occurrence of $c\bar{c}d\bar{d}$ rescattering into a
$c\bar{c}s\bar{s}$ state.

The above decay modes have been studied at \babar.
The $\eta$ is reconstructed in $\gamma \gamma$ or $\pip \pim \piz$
final states and the $\eta^\prime$ in the $\eta(\ra \gamma \gamma)
\pip \pim$ channel.
The $\phi$ is reconstructed in the $K^+K^-$ final state.
Table~\ref{tab:rarejpsi} shows the preliminary
results~\footnote{Unless otherwise stated, charged conjugate modes are
  implied throughout the paper} obtained from the analysis of $50.9$
fb$^{-1}$ of data recorded at the \FourS resonance~\cite{BABrare}.

\begin{table}[t]
\caption{Preliminary branching fraction determinations for rare $B$ decays
  to final states with a $\jpsi$. When the signal yield is not statistically
  significant, a $90 \%$ C.L. upper limit is
  reported.\label{tab:rarejpsi}}
\vspace{0.4cm}
\begin{center}
\begin{tabular}{|l|c|}
\hline
Decay Mode & Branching Fraction \\
\hline
& \\
$B^0 \ra \jpsi \eta(\ra \gamma \gamma)$ &$< 3.0 \times 10^{-5}$
\\
$B^0 \ra \jpsi \eta(\ra \pip \pim \piz)$ &$< 5.2 \times 10^{-5}$
\\
$B^0 \ra \jpsi \eta$(combined) & $< 2.7 \times 10^{-5}$
\\
& \\
\hline 
& \\
$B^0 \ra \jpsi \eta^\prime(\ra \eta(\gamma \gamma) \, \pip \pim)$ & $< 6.4
\times 10^{-5}$ \\
& \\
\hline
& \\
$B^+ \ra \jpsi \phi K^+$ & $\,\,\,(4.4 \pm 1.4 \pm 0.7) \times 10^{-5}$
\\
$B^0 \ra \jpsi \phi K^0$ & $(10.2 \pm 3.8 \pm 1.8) \times 10^{-5}$
\\
$B \ra \jpsi \phi K$ (combined)& $\,\,\,(5.0 \pm 1.3 \pm 0.7) \times 10^{-5}$
\\
& \\
\hline
& \\
$B^0 \ra \jpsi \phi$ &$< 0.95 \times 10^{-5}$
\\ 
& \\
\hline
\end{tabular}
\end{center}
\end{table}

\subsection{Measurement of $B \ra \jpsi \pip \pim$}

In the decay $B \ra \jpsi \pip \pim$, the $\pip \pim$ pair comes from
the $B^0 \ra \jpsi \rho^0(\ra \pip \pim)$ channel or can be produced
in a non-resonant state.
The $B^0 \ra \jpsi \rho^0$ mode is useful for the
measurement of $\sin2\beta$ and possible interference with higher
order diagrams could produce a sizeble
deviation of the branching fraction from the tree level expectation. 
An upper limit on this decay has been set by the CLEO 
Collaboration~\cite{CLEOpipi}.

At \babar, the decay $B^0 \ra \jpsi \pip \pim$ is exclusively
reconstructed and the signal yield is extracted from an unbinned
maximum likelihood fit
to the $\pip \pim$ invariant mass of the selected candidates~\cite{BABpipi}.
The preliminary result obtained from a sample of $51.7$ fb$^{-1}$
of data recorded at the \FourS resonance is 
${\cal B}(B \ra \jpsi \pip \pim)=
(5.0 \pm 0.7 \pm 0.6) \times 10^{-5}$.

\subsection{Measurement of $B \ra \etac K$}

The decay $B^0 \ra \etac K_S$ can be used for a theoretically clean
determination of $\sin2\beta$, in the same way as the ``golden'' mode
$B^0 \ra \jpsi K_S$. 
Previous studies of the neutral and charged decay modes were
performed by the CLEO Collaboration~\cite{CLEOetack}.

At \babar, the decay $B \ra \etac K$ is exclusively reconstructed, 
with the $\etac$
decaying in $K_S K^{\pm} \pi^{\pm}$, $K^+ K^- \piz$ or $K^+ K^- K^+
K^-$ final states~\cite{BABetack}. 
The preliminary results obtained from a data sample of $20.7$
fb$^{-1}$ recorded at the \FourS resonance are ${\cal B}(B^+ \ra \etac K^+)
= (1.50 \pm 0.19 \pm 0.15 \pm 0.46) \times 10^{-3}$ and ${\cal B}(B^0
\ra \etac K^0) = (1.06 \pm 0.28 \pm 0.11 \pm 0.33) \times 10^{-3}$,
where the third error contribution is due to the uncertainty on 
the value of ${\cal B}(\etac \ra K K \pi)$, as reported in the PDG~\cite{PDG}.

\section{Observation of $B^0 \ra \Ds \pi^-$}

One of the methods to determine the angle $\gamma$ of the
unitarity triangle~\cite{BABphyboo} is the measurement of
$\sin(2\beta+\gamma)$ from the time dependent $CP$-asymmetry 
of the decay $B^0 \ra D^+ \pi^-$~\cite{Duni}.
The asymmetry evolution depends on the 
parameter $\lambda_{D\pi} \equiv A(B^0 \ra D^+
\pi^-)/A(B^0 \ra D^- \pi^+)$ which
can be determined from the branching fraction measurement of 
$B^0 \ra \Ds \pi^-$ through the relation:
\begin{equation}
{\cal B}(B^0 \ra \Ds \pi^-) \approx \frac{{\cal B}(B^0 \ra D^- \pi^+)}
{\tan^2 \theta_C} \, \left( \frac{f^2_{D_s}}{f^2_D} \right) 
\, | \lambda_{D\pi}|^2 \, .
\label{eq:lamdipi}
\end{equation}
The above equation is valid in the limit of the tree diagram dominance
for $\Ds \pi^-$ and $D^+ \pi^-$ modes.

At \babar, the decay $B^0 \ra \Ds \pi^-$ is exclusively
reconstructed, with the $\Ds$ decaying in $\phi\pi^+$, 
$\bar{K}^{\ast0} K^+$ or $K_S K^+$ final states. 
From the analysis of a data sample of $56.4$ fb$^{-1}$ recorded at 
the \FourS
resonance, the number of observed signal events is $N_{D_s \pi} = 14.9
\pm 4.1$ with a statistical significance of $3.5 \sigma$. 
The preliminary branching fraction is ${\cal B}(B^0 \ra \Ds \pi^-)
\times {\cal B}(\Ds \ra \phi\pi^+) = (1.11 \pm 0.37 \pm 0.24) \times
10^{-6}$. 
Using the value of ${\cal B}(\Ds \ra \phi\pi^+)$ in the PDG, 
which has a $25 \%$ uncertainty, a branching fraction ${\cal B}(B^0 \ra \Ds
\pi^-)=(3.1 \pm 1.0 \pm 1.0) \times 10^{-5}$ is obtained. 

\section{$D$ meson production in $B^0$ decays}

Inclusive branching fractions of charged and neutral $B$ mesons to
charmed hadrons will help to solve the longstanding {\it $n_c$
  puzzle}~\cite{yamamoto}: 
the mean number of charm quarks per $B$ decay obtained from 
direct counting
does not agree with theoretical estimates based on branching fraction
measurements of semileptonic decays.

The analysis of $D$ meson production in $B^0$ decays at \babar \, is
based on the exclusive reconstruction of one $B$ meson coming from the
decay of the
\FourS ($\equiv B^0_{reco}$) in a semileptonic ($D^\ast l \nu$,
with $l=e,\mu$) or hadronic mode ($D^{(\ast)}\pi^-$,
$D^{(\ast)}\rho^-$, $D^{(\ast)}a_1^-$). 
The recoil system is then analyzed to search for a neutral (charged)
$D$ in the $D^0 \ra K \pi$ ($D^\pm \ra K \pi \pi$) channel. 
The inclusive branching fractions ${\cal B}(B^0 \ra D)$ and ${\cal
  B}(B^0 \ra D^\pm)$ are determined from a fit to the invariant mass
distribution of the selected $D$ candidates.

\begin{table}[t]
\caption{Preliminary branching fraction measurements of flavor-tagged
  $D$ meson production in $B^0$ decays. World data values for $D$
  meson production in $B$ decays are reported for comparison. Here
  ``$B$'' is an admixture of charged and neutral $B$ mesons at
  the \FourS.\label{tab:inclBtoD}}
\vspace{0.4cm}
\begin{center}
\begin{tabular}{|l|c|l|c|}
\hline
\multicolumn{2}{|c|}{\babar\, Measurements} &\multicolumn{2}{|c|}{World Data}   \\
\hline
Decay Mode & Branching Fraction & Decay Mode & Branching Fraction \\
\hline
& & &\\
$\bar{B}^0 \ra D^0$ & $(50.3 \pm 3.0 \pm 3.7 ) \% $ & & \\
$\bar{B}^0 \ra D^+$ & $(32.8 \pm 2.5 \pm 3.5 ) \% $ & & \\
$\bar{B}^0 \ra D^0+D^+$ & $(83.1 \pm 6.4) \%\:\:\:\:\:\:\:\:\:\:\,\, $ & $\bar{B} \ra
D^0+D^+$ & $(78.5 \pm 3.4) \% $~\cite{PDG}\\
$\bar{B}^0 \ra \bar{D}^0$ & $\,\,\,(7.6 \pm 1.7 \pm 1.1 ) \% $ & $\bar{B} \ra \bar{D}^0$ & $\,\,\,(7.3 \pm 3.8) \% $~\cite{delphi}\\
$\bar{B}^0 \ra D^-$ & $\,\,\,(2.7 \pm 1.2 \pm 0.6 ) \% $ & $\bar{B}
\ra D^-$ & $\,\,\,(2.7 \pm 1.7) \% $~\cite{delphi}\\
& & & \\
\hline
\end{tabular}
\end{center}
\end{table}

In the inclusive decays $\bar{B}^0 \ra D^0$ and $\bar{B}^0 \ra D^+$
the charm quark comes directly from the decaying $b$ quark, and the
$D$ meson is said to be of ``right-sign''.
On the contrary, the $D$ meson in the inclusive decays $\bar{B}^0 \ra
\bar{D}^0$ and $\bar{B}^0 \ra D^-$ is said to be of ``wrong-sign''. 
The fraction $w$ of decays with a ``wrong-sign'' $D$  is determined by
comparing the flavor of the $B^0_{reco}$ with that of the $D$, after
correcting for the $B$ mixing probability $\chi_d$:
\begin{equation}
\chi_{obs} = \chi_d + w \times (1-2 \, \chi_d) \, .
\label{eq:wfrac}
\end{equation}
If $\Delta t$ is the time difference between the decays of the two $B$
mesons, events with $|\Delta t| > 2.5 ps$ are discarded.
Indeed, they do not contribute
significantly to $w$ measurement because $\chi_d(|\Delta t| > 2.5 ps) =
1/2$.
The requirement on $|\Delta t|$ increases the sensitivity to $w$, thereby
improving the statistical error.
It improves also the systematic error since the reduced contribution from the 
$B$ mixing.

Preliminary measurements of ${\cal B}(B^0 \ra D)$ and 
${\cal B}(B^0 \ra D^\pm)$ are based on a sample of $30.4$
fb$^{-1}$, while the fractions $w$ are
determined from a sample of $51.1$ fb$^{-1}$, all recorded at the
\FourS resonance. 
These determinations are combined to obtain the first measurements 
of flavor tagged $D^0$ and $D^\pm$
production in $B^0$ decays.
Preliminary results are shown in Table~\ref{tab:inclBtoD}. They agree
with existing measurements of flavor tagged $D$ meson
production in a \FourS environment.
The flavor of the spectator quark in the parent $B$ appears to
have a negligible effect in the production of ``wrong-sign'' $D$
mesons. 
The increase in the central value of the measured branching fractions
goes in the direction of a better agreement with theoretical
predictions. However, in order to solve the {\it $n_c$
  puzzle} other inclusive branching fraction measurements are
needed.


\begin{thebibliography}{99}
\bibitem{NIMbab} \babar \, Collaboration, B.~Aubert {\it et al.},
  \nima{479}, 1-116 (2002). 

\bibitem{PEPII} ``PEP-II -- An Asymmetric $B$ Factory'', Conceptual Design
  Report, SLAC-R-418, LBL-5379 (1993). 

\bibitem{jpsiPRD} \babar \, Collaboration, B.~Aubert {\it et al.},
  \jprd{65}, 032001, (2002). 

\bibitem{L3ref} L3 Collaboration, M.~Acciarri {\it et al.}, \plb{391},
  481, (1997). 

\bibitem{CLEOphiK} CLEO Collaboration, A.~Anastassov {\it et al.},
  \jprl{84}, 1393 (2000). 

\bibitem{BABrare} \babar \, Collaboration, B.~Aubert {\it et al.},
  \babar-CONF-02/06, SLAC-PUB-9166, hep-ex/0203035

\bibitem{CLEOpipi} CLEO Collaboration, M.~Bishai {\it et al.},
  \plb{369}, 186 (1996).

\bibitem{BABpipi} \babar \, Collaboration, B.~Aubert {\it et al.},
  \babar-CONF-02/04, SLAC-PUB-9171, hep-ex/0203034. 

\bibitem{CLEOetack} CLEO Collaboration, K.W.~Edwards {\it et al.},
  \jprl{86}, 30 (2001).

\bibitem{BABetack} \babar \, Collaboration, B.~Aubert {\it et al.},
  \babar-CONF-02/05, SLAC-PUB-9170, hep-ex/0203040. 

\bibitem{PDG} Particle Data Group, D.E.~Groom {\it et al.},
  \epj{15}, 1 (2000).

\bibitem{BABphyboo} ``The \babar\, Physics Book: Physics at An Asymmetric
  $B$-Factory'', P.F.~Harrison and H.R.~Quinn, ed., SLAC-R-504 (1998). 

\bibitem{Duni} I.~Dunietz, \plb{427}, 179 (1998). 

\bibitem{yamamoto} H.~Yamamoto, Proceedings of the 8th International
  Symposium on Heavy Flavor Physics (Heavy Flavors 8), hep-ph/9912308. 

\bibitem{delphi} DELPHI Collaboration, DELPHI 2000-105 CONF 404 (30
  June, 2000), Contributed Paper for ICHEP2000. 

\end{thebibliography}
\end{document}